\documentclass[aps,pra, twocolumn, superscriptaddress,showpacs,showkeys,floatfix]{revtex4-1}
\usepackage{amsmath}
\usepackage{bm}
\usepackage{amssymb}
\usepackage{graphicx}
\usepackage{epstopdf}
\usepackage{epsfig}
\usepackage{color}
\usepackage{hyperref}
\usepackage{float}
\usepackage{braket}
\usepackage{lipsum}

\begin{document}

\title{Phase-locked bi-frequency Raman lasing in a double-$\Lambda$ system}
\author{Hadiseh Alaeian}
\affiliation{5. Physikalisches Institut and Center for Integrated Quantum Science and Technology, University of Stuttgart, Pfaffenwaldring 57, 70569 Stuttgart, Germany}
\author{M. S. Shahriar}
\affiliation{Department of Electrical Engineering and Computer Science, Northwestern University, Evanston, Illinois 60208, USA}
\affiliation{Department of Physics and Astronomy, Northwestern University, Evanston, Illinois 60208, USA}

\date{\today}

\begin{abstract}
We show that it is possible to realize simultaneous Raman lasing at two different frequencies using a double-$\Lambda$ system pumped by a bi-frequency field.  The bi-frequency Raman lasers are phase-locked to one another, and the beat-frequency matches the energy difference between the two meta-stable ground states.  Akin to a conventional Raman laser, the bi-frquency Raman lasers are expected to be subluminal.  As such, these are expected to be highly stable against perturbations in cavity length, and have quantum noise limited linewidths that are far below that of a conventional laser.  Because of these properties, the bi-frequency Raman lasers may find important applications in precision metrology, including atomic interferometry and magnetometry.  The phase-locked Raman laser pair also represent a manifestation of lasing without inversion, albeit in a configuration that produces a pair of non-degenerate lasers simultaneoulsy.  This feature may enable lasing without inversion in frequency regimes not accessible using previous techniques of lasing without inversion.  To elucidate the behavior of this laser pair, we develop an analytical model that describes the stimulated Raman interaction in a double-$\Lambda$ system using an effective 2-level transition. The approximation is valid as long as the excited states adiabatically follow the ground states, as verified by numerical simulations. The effective model is used to identify the optimal operating conditions for the bi-frequency Raman lasing process.  This model may also prove useful  in other potential applications of the  double-$\Lambda$ system, including generation of squeezed light and spatial solitons.

\end{abstract}

\maketitle
\section{Introduction}
When exciting a multilevel atom with more than one coherent field, new phenomena can arise. In such a system, interference between different pathways can lead to suppression of excitations, leading to  a trapped state. In recent years, quantum coherence in multi-level atoms and solid-state emitters has been substantially used for realizing optical bistability~\cite{Wang12, Wang16, Asadpour14, Li16, Li17}, electromagnetic induced transparency (EIT)~\cite{Boller91, Harris97, Marangos98, Fleischhauer00, Wang17}, and subluminal/superluminal light propagation and lasers~\cite{Zhou16, Yablon17}. Moreover, various quantum coherence phenomena in $\Lambda$- and $V$-systems have been employed to modify the linear and non-linear optical characteristics of the medium, and control the temporal and spatial profiles of pulses within that medium~\cite{Kasapi95,Lukin01,Fleischhauer05}.

Broadly speaking, multi-level transitions can be classified into two groups: open-loop and closed-loop. In an open-loop configuration, the transition from any level to another is excited through a single path. In contrast, in a closed-loop system, the transition from one level to another is excited through more than one paths. In an open-loop system, the dynamics is insensitive to the phase of the fields, and is controlled by the amplitudes thereof only. On the other hand, in a closed-loop system the phases of the fields play a critical role~\cite{Buckle86}, thus enriching the range of possible applications. Among different closed-loop schemes, the double-$\Lambda$ systems have drawn a lot of attention recently. These systems consist of a pair of $\Lambda$-configurations with common ground states. The double-$\Lambda$ scheme has been widely used to produce important phenomena, such as coherent population trapping (CPT)~\cite{Zanon05, Donley13}, spin squeezing~\cite{Dantan03}, entanglement~\cite{Zavatta14}, four-wave mixing~\cite{Babin96, Hemmer95, Lü98}, and entangled images~\cite{Boyer08, Miller08}. Unlike conventional $\Lambda$-systems however, many of these important coherent features, like EIT, can only be satisfied for certain relative phases between the laser fields in a double-$\Lambda$ system. This closed-loop phase constraint has substantial effects on the dynamics of the double-$\Lambda$ system, and has been utilized to control the behavior of atoms phase-sensitively~\cite{Xu13,Korsunsky99}.

Due to this unique feature, the double-$\Lambda$ system is expected to play an important role in atom-laser interaction. Therefore, it is useful to develop an easy physical insight and yet accurate description for such systems. In general, the dynamics of a double-$\Lambda$ configuration as a 4-level system can be properly described with 15 real variables, assuming that the total number of atoms is conserved. The time evolution of the level populations and their coherences are determined via the Liouville's equation, with Lindblad terms describing various dephasing and decoherence phenomena. However, due to the complexity of the problem and the large number of variables involved in such a system, the role of various laser fields in controlling the final atomic features are often obscured and drawing a physical insight is hindered. 

On the other hand, the 2-level system has been the canonical prototype for atom-laser interaction, and its physical features have been transparently understood. Therefore, the dynamics of a complicated multi-level system can be more easily visualized if it could be reduced to an effective 2-level one.
Extending the previous work on the $\Lambda$-system~\cite{Shahriar97}, we first propose a generalizable method that reduces a multi-$\Lambda$ system to an effective 2-level atom. Although the numerical results presented in this paper is only for double-$\Lambda$ systems, the method can be easily extended to multiple-$\Lambda$ systems as long as frequency differences of the laser fields in each $\Lambda$-sub system are the same. Using the explicit terms of the effective Hamiltonian the role of the relative phases of the laser fields can be clearly seen. In particular, we study the effect of this phase on controlling the dispersion of the medium and derive the conditions for achieving optical gain for a pair of laser fields in one of the $\Lambda$-sub systems. We show the possibility of simultaneous lasing for these two beams when the pumped medium is inserted inside an optical cavity with a particular length. This dual-beam laser would only be realized when both the relative amplitudes as well as the phases of the optical beams satisfy specific constraints.

This paper is organized as follows; The general theoretical model and comparison between the numerical results of the double-$\Lambda$ system and its effective 2-level model are presented in the second section. The effect of the relative phase and amplitude of the laser fields on the medium dispersion have been investigated in this section as well. In the third section we investigate the bi-laser problem by searching the parameter space to obtain simultaneous optical gain for both transitions of one of the $\Lambda$-sub systems. We derive the required and sufficient conditions for dual-beam lasing. For a particular case that could be realized using $^{87}$Rb, for example, we calculate some of the typical values of the cavity length, laser frequencies, and required optical gain for a moderate quality factor cavity. Finally, section four summarizes the paper and presents the outlook and perspectives for future works.

\section{Theoretical model}
\begin{figure*}
\includegraphics[scale=0.87]{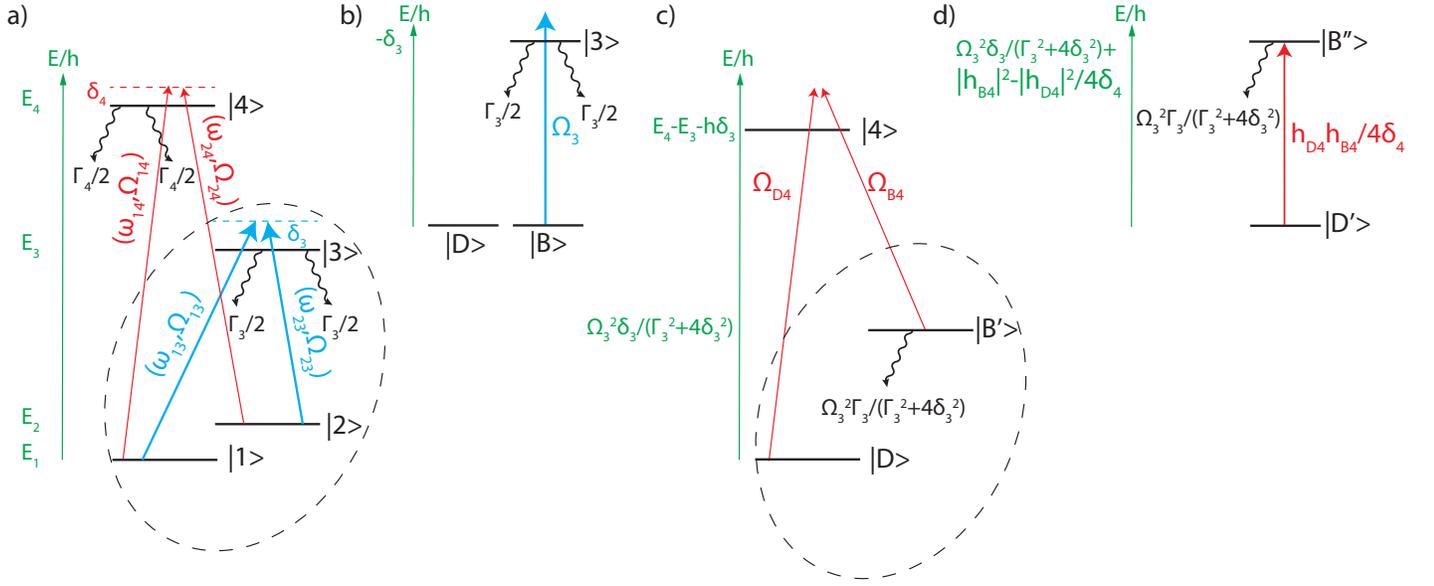}
\caption{\label{Fig1} (a) Energy diagram of a double-$\Lambda$ system and the excitation scheme studied in this work. (b) The energy levels and interaction strength of the pump-$\Lambda$-system encircled in panel (a) in the dark, bright, and 3-level states. (c) Energy levels and transitions in the probe-$\Lambda$ system after replacing the pump-$\Lambda$ system with its equivalent 2-level model. (d) Level diagram and coherent interaction in the ultimate 2-level system equivalent to the original double-$\Lambda$ configuration.}
\end{figure*}

Figure~\ref{Fig1}(a) shows the energy diagram of a typical 4-level, double-$\Lambda$ system investigated in this study. The system is composed of two 3-level sub systems with optical excitations to states $\ket{3}$ and $\ket{4}$, and shared meta-stable ground states $\ket{1}$ and $\ket{2}$. 
The corresponding coherent interactions and decay rates are shown on the figure as well. A planewave laser field with frequency $\omega_{mn}$, wavevector $k_{mn}$, and a constant phase of $\phi^0_{mn}$ driving a coherent interaction between the states $\ket{m}$ and $\ket{n}$ has the following form:

\begin{equation}\label{planewaves}
\vec{E}_{mn}(\vec{r},t)=\frac{\vec{E}_{mn}}{2}~\text{exp}[i(-\omega_{mn}t + \vec{k}_{mn}\cdot \vec{r} + \phi^0_{mn})] + c.c.
\end{equation}

All the couplings are electric-dipole transitions with the Rabi frequency of $\Omega_{mn} = \frac{\vec{M}_{mn}\cdot \vec{E}_{mn}}{\hbar}$, where $\vec{M}_{mn} = \braket{\vec{m}|\hat{p}|\vec{n}}$ is the transition dipole moment between $\ket{m}$ and $\ket{n}$. In this model, it is assumed that there is no electric dipole transition between states $\ket{3}$ and $\ket{4}$, and between states $\ket{1}$ and $\ket{2}$. Furthermore, we assume that each optical field interacts with only one transition, and its couplings to the remaining transitions are either too far detuned or prohibited by polarization selection rules. 

For each transition, the frequency detuning is defined as $\delta_{mn} = \omega_{mn} - (E_m - E_n)/\hbar$.
In each $\Lambda$-system one can define the common detuning $\delta_{3(4)}$ and difference detuning $\Delta_{3(4)}$ as:

\begin{subequations}\label{common-difference detuning}
\begin{align}
\delta_{3(4)} = \frac{\delta_{13(14)} + \delta_{23(24)}}{2}\\
\Delta_{3(4)} = \delta_{13(14)} - \delta_{23(24)}
\end{align}
\end{subequations}

Throughout this paper we use the rotating wave approximation, and work in the transformed basis which is related to the atomic basis via the following equations:
\begin{subequations}\label{rotated frame}
\begin{align}
\ket{\tilde{1}} ={}& \ket{1}\text{exp}[i(\omega_{13}t - \vec{k}_{13}\cdot\vec{r} - \phi^0_{13})]\\
\ket{\tilde{2}} ={}& \ket{2}\text{exp}[i(\omega_{23}t - \vec{k}_{23}\cdot\vec{r} - \phi^0_{23})]\\
\ket{\tilde{3}} ={}& \ket{3}\\
\ket{\tilde{4}} ={}& \ket{4}\text{exp}[-i((\omega_{14}-\omega_{13})t - (\vec{k}_{14} - \vec{k}_{13})\cdot\vec{r} - (\phi^0_{14}-\phi^0_{13}))]
\end{align}
\end{subequations}

The dynamics of the system can be completely described via the Liouville equation for the density matrix. As will be shown later, a steady-state solution without pulsations in the populations of the atomic states only exists if two-photon detunings of both $\Lambda$-sub systems are the same i.e. $\Delta_3 = \Delta_4$. In what follows we further assume that this difference detuning is zero i.e. $\Delta_{3(4)} = 0$ as illustrated in Fig.~\ref{Fig1}(a).

Among various excitation schemes, we are interested in the cases where the transitions to $\ket{3}$ are stronger than the transitions to $\ket{4}$, i.e. $\Omega_{13(23)} \gg \Omega_{14(24)}$. For brevity, we refer to the stronger (weaker) beams as pumps (probes). This assumption allows one to treat the evolution of the probe-$\Lambda$ system as a perturbation.

The Raman excitation for the pump-$\Lambda$ system is better described in the dark and bright states: $\ket{D}$ and $\ket{B}$~\cite{Shahriar97}. The dark state $\ket{D}$ is the superposition of the two meta-stable ground states with zero transition dipole moment to $\ket{3}$. The bright state $\ket{B}$ on the other hand, is orthogonal to $\ket{D}$ with maximized transition dipole moment to $\ket{3}$. In terms of the rotated states of $\ket{\tilde{1}},\ket{\tilde{2}}$, the dark and bright states are defined as:

\begin{subequations}\label{bright-dark definition}
\begin{align}
\ket{D} \equiv \frac{\Omega_{23}\ket{\tilde{1}} - \Omega_{13}\ket{\tilde{2}}}{\Omega_3}\\
\ket{B} \equiv \frac{\Omega_{13}\ket{\tilde{1}} + \Omega_{23}\ket{\tilde{2}}}{\Omega_3}
\end{align}
\end{subequations}

where $\Omega_3 \equiv \sqrt{\Omega^2_{13} + \Omega^2_{23}}$.
Figure~\ref{Fig1}(b) shows the energy levels of this transformed basis and its corresponding transitions. As depicted for zero difference detuning (i.e. $\Delta_3 = 0$), the dark and bright states are degenerate and there is a coherent interaction between the bright and the excited state $\ket{3}$ with Rabi frequency of $\Omega_3$.

After the adiabatic elimination of the excited state $\ket{3}$ in the damped amplitude equation~\cite{Shahriar14}, the pump-$\Lambda$ system can be replaced with an equivalent 2-level model governed by the following Hamiltonian:

\begin{equation}\label{2-level pump}
\hat{\tilde{H}}_{2-level}^{123} = \frac{\hbar}{2}L_3 (2\delta_3 -i\Gamma_3)\ket{B'}\bra{B'}
\end{equation}
where $\ket{B'}$ is the light-shifted version of $\ket{B}$, and $L_3 = \frac{\Omega_3^2}{\Gamma_3^2 + 4\delta_3^2}$ is a factor determined solely by the parameters of the pump-$\Lambda$ system.

This is a non-Hermitian Hamiltonian~\cite{Shahriar14} that accounts for the decay of state $\ket{B'}$ into state $\ket{D}$, and the decay of the coherence between states $\ket{D}$ and $\ket{B'}$. For conservation of the  number of atoms, one still has to add a source term to the population of $\ket{D}$, in formulating the density matrix equation for this effective 2-level system. It should be noted that $\ket{D}$ and $\ket{B'}$ are no longer degenerate in energy, due to the light shift of $\ket{B'}$.

We now consider the addition of the excitation applied to the probe-$\Lambda$ sub-system, as a perturbation. This is illustrated schematically in Fig.~\ref{Fig1}(c). As derived in the Appendix~\ref{app1}, the complex Hamiltonian describing the resulting system, consisting of states $\ket{D},\ket{B'}$ and $\ket{\tilde{4}}$, can be expressed as:
\begin{equation}\label{RW-probe-lambda}
\begin{split}
\hat{\tilde{H}} & = \frac{\hbar}{2} [L_3 (2\delta_3 - i \Gamma_3) \ket{B'}\bra{B'} - (2\delta_4 + i\Gamma_4)\ket{\tilde{4}}\bra{\tilde{4}} \\
& - \frac{1}{\Omega_3}(\Omega_{23}\Omega_{14} - \Omega_{13}\Omega_{24}~e^{-i\Phi_0}) \ket{D}\bra{\tilde{4}} + h.c. \\
& - \frac{1}{\Omega_3}(\Omega_{13}\Omega_{14} + \Omega_{23}\Omega_{24}~e^{-i\Phi_0})\ket{B'}\bra{\tilde{4}} + h.c.]
\end{split}
\end{equation}
where $\Phi_0 = (\phi_{24} - \phi_{23}) - (\phi_{14}-\phi_{13})$ is the closed-loop phase.

Just as before, state $\ket{\tilde{4}}$ can be adiabatically eliminated, using the damped amplitude equations, as shown in the Appendix~\ref{app1}. The resulting Hamiltonian can be expressed as:

\begin{equation}\label{equivalent final 2-level}
\begin{split}
\hat{\tilde{H}}_{2-level}^{1234} &= \frac{\hbar}{2}[\frac{|h_{D4}|^2}{2\delta_4} \ket{D'}\bra{D'}  \\
& + \{ \frac{|h_{B'4}|^2}{2\delta_4} + L_3
(2\delta_3 - i\Gamma_3)\} \ket{B''}\bra{B''} \\
& + \frac{h_{D4} h_{B'4}^*}{2\delta_4} \ket{D'}\bra{B''} + h.c.]
\end{split}
\end{equation}

where $\ket{D'}$ is the light-shifted version of $\ket{D}$, and $\ket{B''}$ is the light-shifted version of $\ket{B'}$. The parameters $h_{D4}$ and $h_{B'4}$ are defined in the Appendix~\ref{app1}, and depend on all Rabi frequencies and the closed-loop phase $\Phi_0$. In the limiting case where $\Phi_0 = 0$ and $\Omega_{23}/\Omega_{13} = \Omega_{24}/\Omega_{14}$, $h_{D4} = 0$. In that case $\ket{D'} = \ket{D}$ and there is no coupling between $\ket{D'}$ and $\ket{B''}$, since the dark state for the pump-$\Lambda$ and probe-$\Lambda$ systems are the same.

The energy diagram for this final 2-level system is shown in Fig.~\ref{Fig1}(d). 
As can be inferred, the interference of the pump and probe beams in the double-$\Lambda$ scheme builds up a coherent interaction between the dark and bright states, whose strength depends on Rabi frequencies of all transitions as well as the closed-loop phase $\Phi_0$. The dependency on $\Phi_0$ is periodic with period $2\pi$. Unlike the conventional $\Lambda$-system, this is a unique feature of such a double-$\Lambda$ configuration, embodying the substantial effect of the laser field phases on the system behavior.

The Hamiltonian in eq.~\ref{equivalent final 2-level} accounts for the decay of state $\ket{B''}$ as well as the dephasing of the coherence between $\ket{D'}$ and $\ket{B''}$. However, in order to conserve the total number of atoms in the system, one must add a source term to the population of state $\ket{D'}$ in order to establish the density matrix equations of motions (i.e. the Liouville equations) for this 2-level system. The Liouville equations can be solved in steady state to find the values of $\rho_{D'D'}, \rho_{B''B''}$, and $\rho_{B''D'} = \rho_{D'B''}^*$. The \textit{approximate} dynamics of the population of state $\ket{\tilde{4}}$ as well as coherence between this state and the meta-stable ground states can be determined using of the relations between the relevant amplitudes established during the process of adiabatic elimination. As shown in Appendix~\ref{app2}, using this procedure we can write:

\begin{equation}\label{approximated coherent terms 4}
\begin{split}
\tilde{\rho}_{14} =\frac{\Omega_4}{2\delta_4 - i\Gamma_4}
\{& (-\cos{\theta_3} \sin{\theta_4} + \sin{\theta_3} \cos{\theta_4} e^{-i\Phi_0})\times \\
& (\cos{\theta_3}~\rho_{DD} + \sin{\theta_3}~\rho_{BD}) - \\
& (\sin{\theta_3} \sin{\theta_4} + \cos{\theta_3} \cos{\theta_4} e^{-i\Phi_0})\times \\
& (\cos{\theta_3}~\rho_{DB} + \sin{\theta_3}~\rho_{BB})
\}
\end{split}
\end{equation}
where $\cos{\theta_{3(4)}} \equiv \Omega_{23(24)}/\Omega_{3(4)}$.

The ultimate equivalent 2-level system presented in eq.~\ref{equivalent final 2-level} reduces the number of unknowns from 15 real variables to 3, hence making the calculations very fast and efficient. Moreover, the equivalent system provides an insight into the mutual interaction between the pumps, probes, and their interference. It is worth mentioning that the reduction scheme described here can be easily extended to multiple-$\Lambda$ systems as long as the difference detuning of all $\Lambda$-sub systems (i.e. $\Delta_i$) are the same.

\begin{figure*}
\includegraphics[scale=0.8]{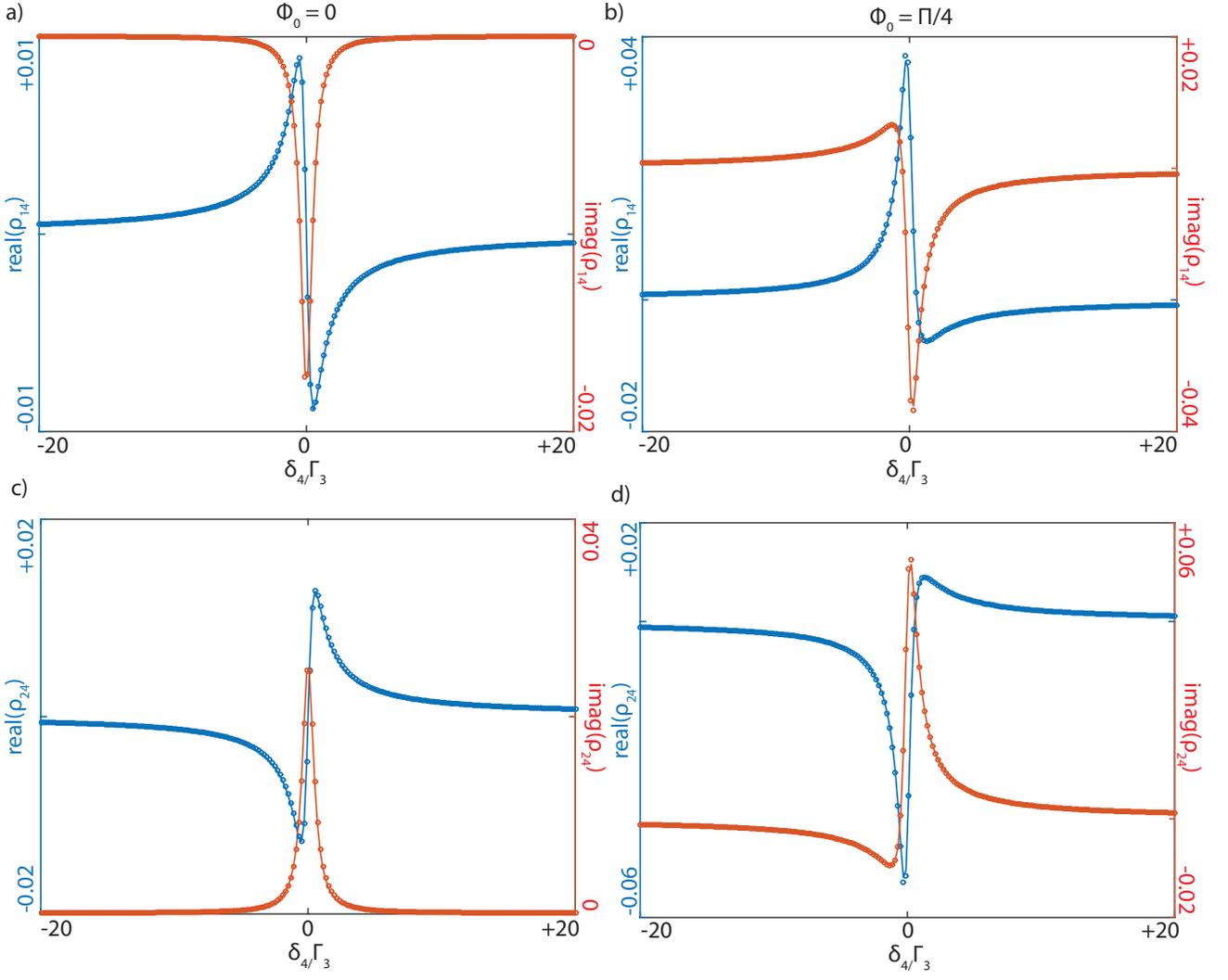}
\caption{\label{Fig2} Comparison between the exact (solid lines) and approximate solutions using the effective 2-level system (dots) of the double-$\Lambda$ system for coherent transitions to the 4$^{th}$-level as a function of detuning $\delta_4$, when $\Omega_{23} = 10\Gamma_3, \Omega_{23} = 7\Gamma_3, \Omega_{14}=\Gamma_3/5,\Omega_{24} = \Gamma_3/2, \delta_3 = \Gamma_3$. (a),(b) show the results for $\rho_{14}$ at two different closed-loop phases of $\Phi_0 = 0$ and $\Phi_0 = \pi/4$, respectively. (c),(d) show the results for $\rho_{24}$ with the same $\Phi_0$ as for (a) and (b). In each panel the real parts are shown in blue while the imaginary parts are denoted in red.}
\end{figure*}

To investigate the validity and accuracy of this approximation, in Fig.~\ref{Fig2} we compare the behavior of $\tilde{\rho}_{14}$ and $\tilde{\rho}_{24}$ as a function of $\delta_4$ for different closed-loop phases. The calculations are done for an ideal 4-level system, with $\Omega_{13}=10\Gamma_3, \Omega_{23}=7\Gamma_3$, $\delta_3 = \Gamma_3$, $\Omega_{14} = \Gamma_3/5,\Omega_{24}=\Gamma_3/2$. Here, we have used $\Gamma_4 = 1.05\Gamma_3$ reflecting the ratio of decay rates of the $5~^2 P_{3/2}$ and $5~^2 P_{1/2}$ manifolds in $^{87}$Rb~\cite{Steck}, which is a possible atom for realizing such a double-$\Lambda$ system.

In all cases the real and imaginary parts are denoted in blue and red, respectively. Moreover, in each panel the solid lines show the exact values calculated via the complete $4\times 4$-density matrix while the dots show the approximated results obtained from eq.~\ref{equivalent final 2-level} and ~\ref{approximated coherent terms 4}. A very good agreement between the exact and the approximated results can be observed for both coherent terms at various $\Phi_0$.

We now consider, analytically, using the reduced 2-level model, certain special cases. Consider first the case where the relative intensities of the two legs are the same for both pump- and probe-$\Lambda$ systems. Specifically, we assume $\Omega_{14}\Omega_{23} = \Omega_{13}\Omega_{24} = \Omega^2$. Using this condition, the coherent interaction between the dark and 4$^{th}$ level in eq.~\ref{RW-probe-lambda} would be further simplified to $\frac{\Omega^2}{\Omega_3}(e^{-i\Phi_0} - 1)$. When $\Phi_0 = 0$ the interaction vanishes and $\ket{D}$ becomes decoupled from $\ket{\tilde{4}}$. In other words when this condition is satisfied both $\Lambda$-sub systems share the same dark state. Moreover, according to eq.~\ref{equivalent final 2-level} the coherent interaction between dark and bright state vanishes as well, hence a coherent population trapping (CPT) happens for the dark state which is independent of individual values of Rabi frequencies. On the other hand, when $\Phi_0 = \pi$ the interaction between the dark state and the 4$^{th}$-level is maximized.

A complementary situation happens when $\Omega_{13}\Omega_{14} = \Omega_{23}\Omega_{24} = \Omega^2$. In that case, as eq.~\ref{RW-probe-lambda} implies, the coherent interaction strength between the bright state and the 4$^{th}$-level is $-\frac{\Omega^2}{\Omega_3}(1+e^{-i\Phi_0})$. In contrast to the previous case, here for $\Phi_0 = 0$, the interaction is maximized and for $\Phi_0 = \pi$ the $\ket{B'}$ and $\ket{\tilde{4}}$ states get decoupled. Also, as can be seen in eq.~\ref{equivalent final 2-level}, this decoupled condition leads to a zero coherent interaction between the dark and bright states. Therefore, similar to the previous case, the bright state gets completely decoupled from the dark state.

These two special conditions can be simultaneously satisfied if both of the pump beams and both of the probe beams have the same strength. In terms of the effective Rabi frequencies we have $\Omega_{13} = \Omega_{23}=\Omega_3/\sqrt{2}$ and $\Omega_{14} = \Omega_{24} = \Omega_4/\sqrt{2}$ and the effective Hamiltonian of eq.~\ref{equivalent final 2-level} would be simplified as:

\begin{equation}\label{equal beam Hamiltonian}
\begin{split}
\hat{\tilde{H}}_{2-level}^{1234} &= \frac{\hbar}{2} [\frac{\Omega_4^2}{2\delta_4}\cos{\Phi_0} + L_3 (2\delta_3 - i\Gamma_3) \ket{B''}\bra{B''} \\
& + i\frac{\Omega_4^2}{4\delta_4}\sin{\Phi_0} \ket{D'}\bra{B''} + h.c.]
\end{split}
\end{equation}

This Hamiltonian describes the dynamics of a 2-level system with $\omega_1 = 0$ , $\omega_2 = (\Omega_4^2 \cos{\Phi_0} /4\delta_4  + L_3 \delta_3)$, and the effective Rabi frequency of $\Omega_{eq} = \Omega_4^2/4\delta_4 \sin{\Phi_0}$. Moreover the bright state decays to the dark state with an effective population decay rate of $\Gamma_{eq} = L_3\Gamma_3$.

The terms in the Hamiltonian of eq.~\ref{equal beam Hamiltonian} explicitly show the effect of the closed-loop phase in modulating the strength of the coherent interaction and the energy gap between the dark and bright states. Both the coupling strength and the energy gap are periodic in $\Phi_0$ with a $\pi/2$ phase shift. While the decay rate and the energy offset of the bright state are solely determined with the pump beams, the probes determine the strength of the coherent interaction between the dark and bright states and the energy gap modulation. 

For $\Phi_0 = 0,\pi$ the coupling between $\ket{D'}$ and $\ket{B''}$ vanishes and these two states become totally decoupled. In other words, $\ket{D'}$ is a trapped state in this configuration and EIT occurs for both excited states, namely $\ket{3}$ and $\ket{4}$.

On the other hand if $\Phi_0 = \pi/2,3\pi/2$ the coupling between the dark and bright states is maximized. Therefore, a coherent interaction is built up between these two states whose strength is solely dependent on the probe-$\Lambda$ system. Specifically, it is proportional to the Rabi frequency of the original atomic states (i.e. $\Omega_{4}$) and decreases as the detuning $\delta_4$ increases. This interaction would lead to population exchange between $\ket{D'}$ and $\ket{B''}$ and consequently populates the excited atomic states $\ket{3}$ and $\ket{4}$, which leads to non-zero coherent terms for $\rho_{14}(24)$ and induces a polarizability for these transitions.

Solving for the steady-state of the system described by eq.~\ref{equal beam Hamiltonian}, we get the following expressions for the effective 2-level density matrix elements:
\begin{subequations}
\begin{align}
\rho_{B''B''} = \frac{\Omega_4^4 \sin^2{\Phi_0}}{2\Omega_4^4(1+\cos^2{\Phi_0}) + 16L_3\Omega_3^2 \delta_4^2+ 32 L_3 \Omega_4^2 \delta_3\delta_4\cos{\Phi_0}}\\
\rho_{D'B''} = -\frac{4\Omega_4^2 \delta_4 \sin{\Phi_0}L_3(\Gamma_3 +i2\delta_3)+i\Omega_4^4 \sin{2\Phi_0}}{2\Omega_4^4(1+\cos^2{\Phi_0}) + 16L_3\Omega_3^2 \delta_4^2+ 32 L_3 \Omega_4^2 \delta_3\delta_4\cos{\Phi_0}}
\end{align}
\end{subequations}

It is clear that the coherence between the dark and bright states contains three types of terms: 1) terms that are only related to $\Omega_4$ (self-terms), 2) those only related to $\Omega_3$ (cross-terms), and 3) terms related to both (mutual terms). This is an important feature of this double-$\Lambda$ system, showing how the pumps and the probes can be selected properly to maximize independently the non-linearities, while suppressing single photon absorptions.

As the coherent interaction strength is completely tunbale with $\Phi_0$, this closed-loop phase can be utilized further to tune the polarizablity of the medium. Combined with Maxwell's equations, the propagation of the pumps and probes can be studied in such a double-$\Lambda$ configuration. In the next section we study the parameter space of achievable polarizabilities in this system and investigate the possibility of simultaneous phase-locked lasing at two different frequencies in a single cavity.

\section{bi-frequency Raman lasing in double-$\Lambda$ configuration}

Assume that all the beams are planewaves and their profiles do not change as they propagate through the medium. In other words, we ignore the effect of the slowly varying envelopes for the first order analysis here. The current analysis can be easily extended by considering the effect of time and position dependent slowly varying envelopes in such a medium.

Depending on various parameters, the probe beams that excite the coherence represented by $\rho_{14(24)}$ can experience optical gain or loss. Upon satisfying proper criteria, both probe beams can experience optical gain as they propagate through the medium. Therefore, in an optical cavity having resonant modes at both probe frequencies, a double-beam laser can be realized if enough optical gain is available for both probe beams.  In this section we derive the proper conditions for having a self-consistent solution for four beams inside the polarizable medium and investigate the possibility of having a bi-frequency laser in a suitably-designed ring-cavity.

Since the pump beams are assumed to be much stronger than the probes, the un-depleted approximation can be employed for both transitions to state $\ket{3}$, hence ignoring any modifications to $\Omega_{13}$ and $\Omega_{23}$.  From the results of the previous section, we can describe the induced polarizability at the probe frequencies in terms of the coherence terms of $\rho_{14}$ and $\rho_{24}$. By inserting these terms back into the wave equation, we get the following equations describing the propagation of the probes within the medium:

\begin{subequations}\label{probe-propagation}
\begin{align}
((\frac{\omega_{14}}{c})^2 - k_{14}^2)\Omega_{14} &  = -\frac{2\mu_0\omega_{14}^2}{\hbar}N|M_{14}|^2\tilde{\rho}_{14}\\
((\frac{\omega_{24}}{c})^2 - k_{24}^2)\Omega_{24} &= 
-\frac{2\mu_0\omega_{24}^2}{\hbar}N|M_{24}|\tilde{\rho}_{24}e^{+i\Phi_0(z)}
\end{align}
\end{subequations}
where $N$ is the atomic density, and $\mu_0$ is the magnetic permeability.

A self-consistent solution for all four beams could be obtained if the wavevectors satisfy a phase matching condition: $(k_{24} - k_{14}) = (k_{23} - k_{13})$. Moreover, the wevevectors of the probes get modified as the they propagate through the pumped medium as: 

\begin{equation}\label{modified k14-k24}
k_{14,24} = \frac{\omega_{14,24}}{c}(1 + \xi_{14,24})
\end{equation}

Combined with the frequency resonance condition, eq.~\ref{modified k14-k24} leads to $\omega_{14}\xi_{14} = \omega_{24}\xi_{24}$. Plugging eq.~\ref{modified k14-k24} in eq.~\ref{probe-propagation}, we find that $\xi_{14,24}$ are related to the atomic parameters and the induced coherence terms as follows: 

\begin{subequations}\label{xi_14,24 equations}
\begin{align}
\xi_{14} & = \frac{\mu_0 N |M_{14}|^2 c^2}{\hbar \Omega_{14}}\tilde{\rho}_{14}\\
\xi_{24} & = \frac{\mu_0 N |M_{24}|^2 c^2}{\hbar \Omega_{24}}\tilde{\rho}_{24}e^{+i((\phi_{24}^0 - \phi_{23}^0)-(\phi_{23}^0 - \phi_{13}^0))}
\end{align}
\end{subequations}

To have a sustainable oscillation from these two beams in a ring cavity with length $L_c$, the cavity resonance condition of $(k_{14}-k_{24})L_c = 2\pi m$ needs to be satisfied.  Taking into account the condition that  $\omega_{14}\xi_{14} = \omega_{24}\xi_{24}$, and eq. ~\ref{modified k14-k24}, it is clear that the cavity length $L_c$ depends primarily on the energy difference between the meta-stable ground states according to the following equation:

\begin{eqnarray}\label{cavity length}
L_c = \frac{2m\pi \hbar c}{E_2 - E_1}
\end{eqnarray}

The second condition for realizing a lasing mode mandates that the optical gain at both probe frequencies be large enough to overcome all the losses and de-coherence phenomena inside the cavity. As ground state energies are often close together (e.g., the hyperfine splitting in the ground state of $^{87}$Rb is 6.8 GHz, which is nearly six orders of magnitude smaller than the optical transition frequencies), it is fair to assume that the cavity quality factor is almost the same for both probes. For a laser cavity with an output coupler mirror transmittivity of $T$, the imaginary parts of $\xi_{14,24}$ must satisfy the following gain-loss balance condition:

\begin{equation}\label{laser gain condition}
\omega_{14}\xi_{14}'' = \omega_{24}\xi_{24}'' = \frac{T.c}{2L_c} = \frac{T(E_2 - E_1)}{4m\pi \hbar}
\end{equation}

For each cavity length determined by $m$, there is a unique amount of the optical gain to satisfy the lasing condition for both probe beams in the cavity. Just as for the cavity length, this optical gain is determined primarily by the energy gap between the meta-stable ground states.

\begin{figure*}
\includegraphics[scale=0.8]{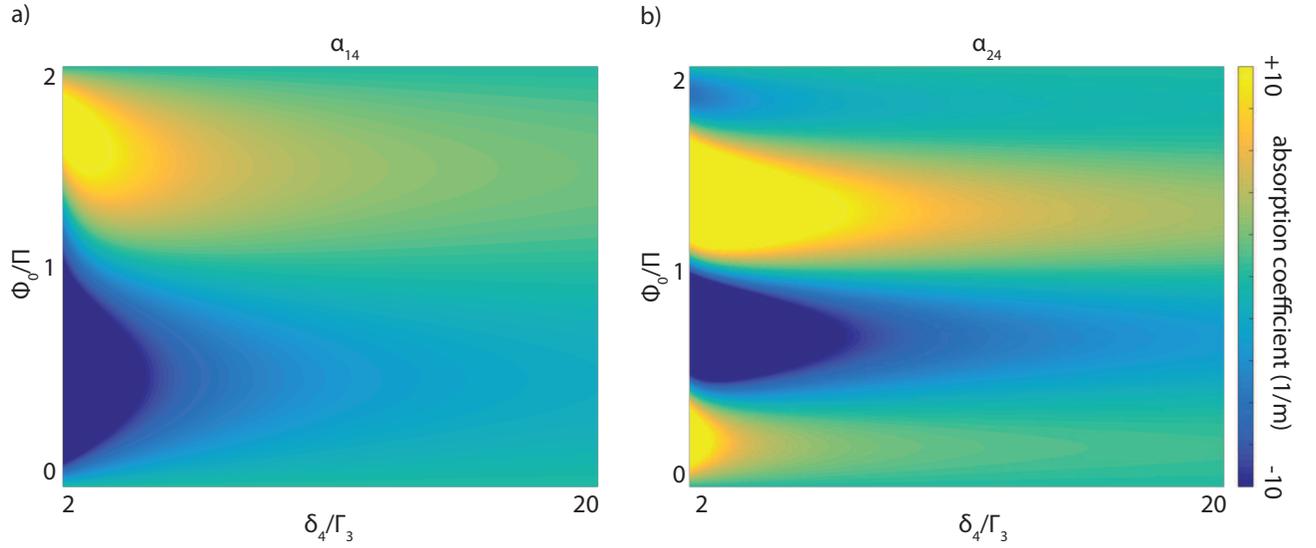}
\caption{\label{Fig3} variation of the absorption coefficient (a) $\alpha_{14}$, (b) $\alpha_{24}$ as a function of difference detuning $\delta_4$ and closed-loop phase ($\Phi_0$). The pump beams are assumed to be equal with $\Omega_3 = 10\Gamma_3$, and the probes have the same intensity of $\Omega_4 = \Gamma_3$. The difference detuning on transitions to $\ket{3}$ is assumed to be $\delta_3 = 10\Gamma_3$, and $\Delta=0$. The particle density is $N=10^{15} m^{-3}$.}
\end{figure*}

Figure~\ref{Fig3}(a) and (b) show the variation of the absorption coefficient $\alpha_{14(24)}=\xi''_{14(24)}\omega_{14(24)}/c$ as a function of probe-$\Lambda$ system detuning ($\delta_4$) and the closed-loop phase ($\Phi_0$) where the atom density is $N = 10^{15}~m^{-3}$. As can be seen these parameters substantially vary in a wide range, making the beams get attenuated (negative regions in blue color) or grow (positive regions in yellow color) as they propagate through the cavity. For lasing to occur, the parameters should be chosen in such a way that both beams experience the same amount of gain (i.e. $\alpha\geq 0$ (yellow regions)). Here both pumps are assumed to have the same strength of $\Omega_3 = 10\Gamma_3$ and the common detuning of $\delta_3 = 10\Gamma_3$. The probes are also assumed to have the same rate of $\Omega_4 = \Gamma_3$.

\begin{figure*}
\includegraphics[scale=0.8]{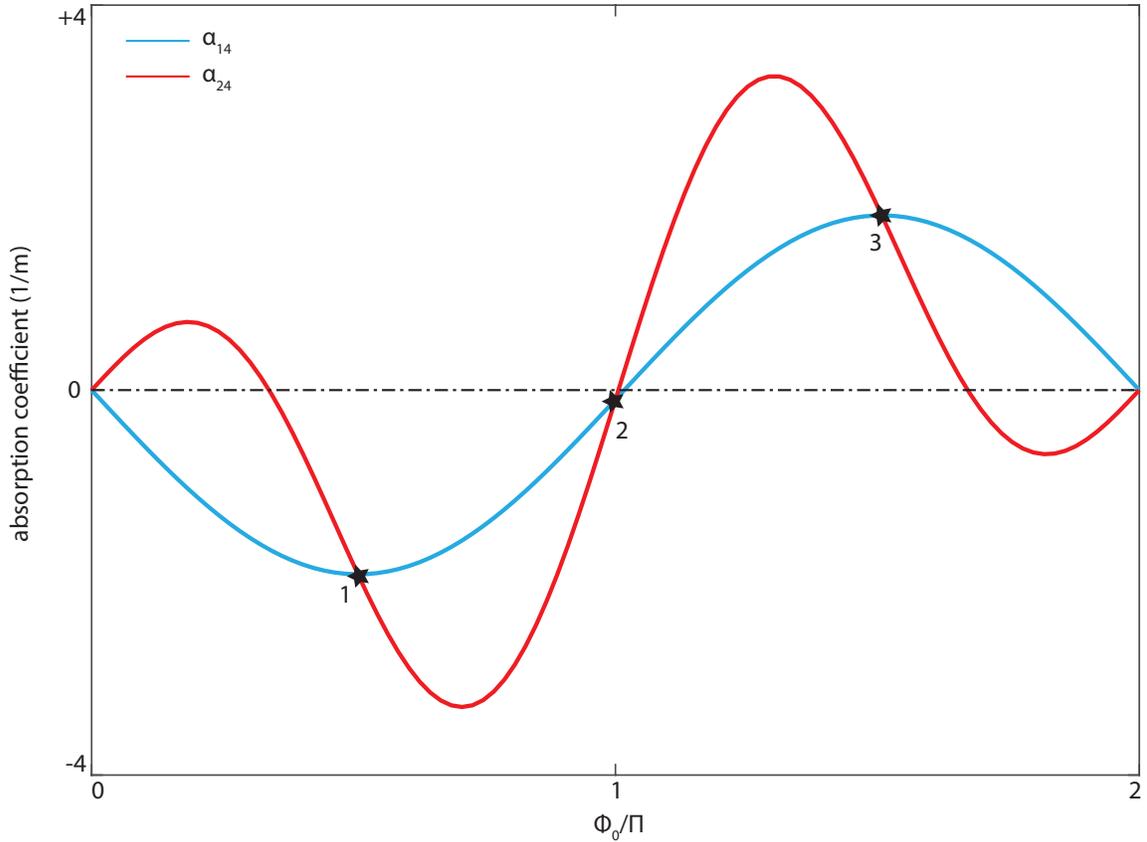}
\caption{\label{Fig4} Variation of the absorption coefficients $\alpha_{14}$ (blue line) and $\alpha_{24}$ (red line) as a function of closed-loop phase. All the parameters are the same as in Fig.~\ref{Fig3} and the detuning for 4$^{th}$ level excitation is assumed to be $\delta_4 = 20\Gamma_3$. The black dashed line shows zero loss/gain condition. The black stars show the points where both transitions experience the same amount of attenuation/amplification. The lasing calculations have been done for $P_3$ where both transitions to $\ket{4}$ experience the same amount of gain.}
\end{figure*}

Figure~\ref{Fig4} shows the variation of the absorption coefficients $\alpha_{14}$ and $\alpha_{24}$ for both transitions to the 4$^{th}$ level at a fixed detuning value of $\delta_4 = 20\Gamma_3$ as a function of $\Phi_0$. As can be seen the polarizability on both transitions varies substantially as a function of this phase and the beams can experience different amounts of gain or loss. In particular there are three points (denoted with black stars) where both beams experience the same effect. The region of our interest is the section where both transitions experience gain as they propagate through the gas. This corresponds to a region with both $\alpha$ being above the black dashed line. The point where they both experience the same gain is denoted as point 3. At this point with $\Phi_0 = 3\pi/2$ the gain experienced by both beams is $\approx 1.8 ~ m^{-1}$. Note that a fixed $\Phi_0$ satisfying the lasing condition implies that phases of the two laser beams must follow a certain relationship, set by the phases of the pump beams.

For a cavity with length of $L=4.38 ~ cm$ (corresponding to $m=1$ in eq.~\ref{cavity length}) this would end up being a per-pass gain of $0.08$, which is large enough to overcome the losses of a ring cavity with the transmittivity of $T\approx 16\%$ for the output coupler mirror.   Once a concrete scheme is adopted for realizing this process (e.g. $^{87}$Rb atoms in a vapor cell), the transition matrix elements would be known, thereby making it possible to determine the values of the electric fields, and hence the intensities, for each laser, since the value of the Rabi frequency is established from the preceding discussions.

\section{conclusion and outlook}
In this paper we have proposed a bi-frequency Raman laser in a double-$\Lambda$ configuration. Unlike conventional Raman lasers the output beams of such a laser are two phase-locked beams separated by a typical value of a few GHz, corresponding to the frequency separation of the meta-stable ground states.  Due to the sensitivity of the gain value to the closed-loop phase, the output modes of the laser are phase-locked and are directly determined via the optical pump phases.

Furthermore, we have described a systematic scheme that produces an equivalent 2-level model for the 4-level system. The equivalent model explicitly shows how each set of pumps and probes, and the closed-loop phase play roles in controlling the final states of the quantum emitter. The generalization of the procedure for multi-$\Lambda$ systems is straight forward as long as all $\Lambda$-sub systems have the same frequency detuning. By analytically solving for the steady-state of the density matrix, we have identified explicitly the contribution of the closed-loop phase, the probe detuning, and the pump Rabi frequencies in controlling the linear susceptibilities for the probes in the double-$\Lambda$ system. 

For devices such as Raman atomic interferometers and CPT clocks, it is necessary to realize a pair of laser frequencies that are phase-coherent with each other, while differing in frequency by the ground state hyperfine splitting of an alkali atom, such as $^87$Rb.  The bi-frequency laser obtained via the proposed scheme uses a pair of such lasers as pumps, and creates another such pair.  However, the bi-frequency laser pair may have properties that are better suited for these applications than the original pump lasers.   It is well known that for these applications, it is important to ensure that the absolute frequency of each laser is as narrow as possible, in order to suppress fluctuations in light-shift.  Recently, we have shown ~\cite{Scheuer15, Zhou16, Yablon17} that a Raman laser acts as a subluminal laser, with a quantum noise limited linewidth (Schwalow-Townes Linewidth: STL) that  is expected to be narrower than that of a conventional laser by a factor equaling the square of the group index.  In reference ~\cite{Yablon17} the observed group index was ~663, with an expected STL of ~1.2 micro-Hz. Since the bi-frequency lasers described here are fundamentally Raman lasers, it is expected that these lasers would also have group indices that are substantially larger than unity, which in turn would imply very small STLs. Of course, the actual group indices for the bi-frequency lasers would depend on the details of the actual atomic transitions employed, the cavity parameters, and the pump powers.  Investigations are in progress to quantify this feature of a bi-frequency laser employing $^{87}$Rb atoms, taking into account non-idealities due to presence of additional energy levels.

The bi-frequency lasing described here is similar to lasers without inversion (LWI), which had been investigated extensively in the past~\cite{Harris89,Imamoglu91,Schully94,Bhatia01}.  However, unlike the conventional LWIs, the bi-frequency laser suggested in this work produces two non-degenerate lasers simultaneously.  These may prove to be easier to implement experimentally, and enable realization of lasers at frequencies for which creation of population inversion has not shown to be possible with existing technologies.  In order to use the bi-frequency lasing process for this goal, it would be necessary to identify a suitable quantum system for which the probe-lambda transition is at a frequency high enough so that a conventional laser does not exist at that frequency, while the pump-lambda transition is at a frequency for which high-power lasers exist.  The resulting bi-frequency laser would transfer energy from the low-frequency pump lasers. Of course, in the model shown in this paper, we have not considered depletion of the pumps, assuming that the power in the bi-frequency laser (probe-lambda) is very low compared to that of the pump.  However, if the mean frequency of the pump-lambda system is much lower than that of the probe-lambda system, this approximation is not a suitable one, and it is necessary to consider a more comprehensive model where the pump depletion is taken into account.  Investigations are underway to identify four-level systems that can be used in this manner to transfer energy from a low-frequency laser to a very high-frequency bi-laser, taking into account pump depletion

As a follow up work, it would be important to extend the study to investigate the phase-sensitive non-linear susceptibility of such a double-$\Lambda$ system, in order to study spatial solitons and their dispersive features. Moreover, it would be useful to utilize the powerful, effective 2-level model for further quantum electrodynamical studies in the context of the Jaynes-Cummings or the Travis-Cummings models.

\appendix
\section{Hamiltonians of the probe-$\Lambda$ system}\label{app1}

In this appendix we present the detailed calculations of the probe-$\Lambda$ Hamiltonian starting from $\ket{D},\ket{B'}, \ket{4}$ states and derive the free and interaction Hamiltonians. Furthermore, we present the required intermediate steps to derive the final effective 2-level Hamiltonian of eq.~\ref{equivalent final 2-level} in the main text.

The free Hamiltonian of the probe-$\Lambda$ sub-system shown in the Fig.~\ref{Fig1}(c) is given via the following equation:

\begin{align}\label{free-Hamiltonian-4level}
\hat{\tilde{H}}_{f} = \frac{\hbar}{2}[L_3 (2\delta_3 - i\Gamma_3) \ket{B'}\bra{B'} + 
2(\omega_4 - \omega_3 - \delta_3)\ket{\tilde{4}}\bra{\tilde{4}}] 
\end{align}

After expressing the dark and bright states in terms of the meta-stable ground states, one can determine the interaction Hamiltonian as give in eq.~\ref{RWA-probe-lambda}.

\begin{widetext} \label{RWA-probe-lambda}
\begin{equation}
\begin{split}
\hat{\tilde{H}}_{int}  = \frac{\hbar}{2} [&\{-\frac{\Omega_{23}\Omega_{14}}{\Omega_3}e^{+i((\omega_{14} - \omega_{13})t-(\phi_{14} - \phi_{13}))} + \frac{\Omega_{13}\Omega_{24}}{\Omega_3}e^{+i((\omega_{24} - \omega_{23})t-(\phi_{24} - \phi_{23}))} \} \ket{D}\bra{4} + h.c.\\
& \{ - \frac{\Omega_{13}\Omega_{14}}{\Omega_3}e^{+i((\omega_{14} - \omega_{13})t-(\phi_{14} - \phi_{13}))} + 
\frac{\Omega_{23}\Omega_{24}}{\Omega_3}e^{+i((\omega_{24} - \omega_{23})t-(\phi_{24} - \phi_{23}))} \} \ket{B'}\bra{4} + h.c.]
\end{split}
\end{equation}
\end{widetext}

From this Hamiltonian it is clear that a unique rotated frame can only exist if the frequencies of the pump and probe beams satisfy $(\omega_{13}-\omega_{23}) = (\omega_{14}-\omega_{14})$. This indeed is the frequency resonance condition as mentioned in the main text.

By rotating $\ket{4}$ to $\ket{\tilde{4}}$ as defined in eq.~\ref{rotated frame}, the Hamiltonian could be simplified to:

\begin{widetext}
\begin{equation}
\begin{split}
\hat{\tilde{H}}_{total}^{RW}  = \frac{\hbar}{2} &[L_3 (2\delta_3 - i \Gamma_3) \ket{B'}\bra{B'} - (2\delta_4 + i\Gamma_4)\ket{\tilde{4}}\bra{\tilde{4}} \\
& - \frac{1}{\Omega_3}\{(\Omega_{23}\Omega_{14} - \Omega_{13}\Omega_{24}~e^{-i\Phi_0}) \ket{D}\bra{\tilde{4}} 
+ (\Omega_{13}\Omega_{14} + \Omega_{23}\Omega_{24}~e^{-i\Phi_0})\ket{B'}\bra{\tilde{4}} + h.c.\}]
\end{split}
\end{equation}
\end{widetext}
which is eq.~\ref{RW-probe-lambda} in the main text.

This 3-level system can be further reduced to an effective 2-level system if the change rate of the 4$^{th}$-level is slower compared to the ground states, so that its dynamics can be adiabatically eliminated and expressed in terms of the lower levels as:

\begin{equation}\label{adiabatic elimination of 4}
\begin{split}
\tilde{c}_4 & = A_D c_D + A_B' c_B'\\
A_D & = \frac{1}{2\delta_4 + i\Gamma_4}(-\frac{\Omega_{23}\Omega_{14}}{\Omega_3} + \frac{\Omega_{13}\Omega_{24}}{\Omega_3}e^{+i\Phi_0}) = \frac{h^*_{D4}}{2\delta_4 + i\Gamma_4}\\
A_B' & = \frac{1}{2\delta_4 + i\Gamma_4}(-\frac{\Omega_{13}\Omega_{14}}{\Omega_3} - \frac{\Omega_{23}\Omega_{24}}{\Omega_3}e^{+i\Phi_0}) = \frac{h^*_{B'4}}{2\delta_4 + i\Gamma_4}
\end{split}
\end{equation}

Substituting these expressions in the Hamiltonian of eq.~\ref{RW-probe-lambda} and assuming $\delta_4 \gg \Gamma_4$ so that the effect of $\Gamma_4$ can be ignored, one would obtain the effective 2-level Hamiltonian of eq.~\ref{equivalent final 2-level}.

\section{Coherence of probe-$\Lambda$ system}\label{app2}
By solving the final 2-level Hamiltonian of eq.~\ref{equivalent final 2-level}, the population of the dark and bright states (i.e. $\rho_{D'D'},\rho_{B''B''}$) and the coherence between these two states (i.e. $\rho_{D'B''}$) can be determined uniquely. Ignoring the small modifications of the dark and bright states due to the adiabatic eliminations of the excited states $\ket{\tilde{3}}$ and $\ket{\tilde{4}}$, the states $\ket{D'}$ and $\ket{B''}$ can be replaced by $\ket{D}$ and $\ket{B}$, respectively. Therefore, the populations and the coherence between ground states can be determined using eq.~\ref{bright-dark definition} as:

\begin{equation}\label{ground state coherence}
\begin{split}
\tilde{\rho}_{11} & = \bra{\tilde{1}}\hat{\rho}\ket{\tilde{1}} = \frac{(\Omega_{23}\bra{D} + \Omega_{13}\bra{B}) \hat{\rho}(\Omega_{23}\ket{D} + \Omega_{13}\ket{B})}{\Omega_3^2}\\ 
& = \cos^2{\theta_3}~\rho_{DD} + \frac{\sin{2\theta_3}}{2}(\rho_{DB} + \rho_{BD}) + \sin^2 {\theta_3}~\rho_{BB}\\
\tilde{\rho}_{12} & = \cos^2{\theta_3}~ \rho_{DB} + \frac{\sin{2\theta_3}}{2}(\rho_{BB} - \rho_{DD}) - \sin^2{\theta_3}~\rho_{BD}
\end{split}
\end{equation}
where $\cos{\theta_3} \equiv \Omega_{23}/\Omega_3$ as defined in the main text.

The coherence between the ground states and the dark and bright states can be determined similarly. For example for the state $\ket{\tilde{1}}$ one has:

\begin{equation}\label{ground state DB-coherence}
\begin{split}
\tilde{\rho}_{1D} & = \bra{\tilde{1}}\hat{\rho}\ket{D} = \frac{(\Omega_{23}\bra{D} + \Omega_{13}\bra{B}) \hat{\rho}\ket{D}}{\Omega_3}\\ 
& = \cos{\theta_3} ~\rho_{DD} + \sin{\theta_3}~\rho_{BD}\\
\tilde{\rho}_{1B} & = \bra{\tilde{1}}\hat{\rho}\ket{B} = \frac{(\Omega_{23}\bra{D} + \Omega_{13}\bra{B}) \hat{\rho}\ket{B}}{\Omega_3}\\ 
& = \cos{\theta_3}~\rho_{DB} + \sin{\theta_3}~\rho_{BB}
\end{split}
\end{equation}

To find the population of the excited states $\ket{\tilde{3}}$ and $\ket{\tilde{4}}$, and their corresponding coherence with other states, one can use the adiabatic elimination equations of amplitudes. Since here we are only interested in the probe-$\Lambda$ system we derive the corresponding relations for this system's transitions.

According to eq.~\ref{adiabatic elimination of 4} the \textit{approximate} coherence between state $\ket{\tilde{4}}$ and the ground states can be determined as:

\begin{equation}\label{4 state 12-coherence}
\begin{split}
\tilde{\rho}_{14} & \approx \tilde{c}_1\tilde{c}_4^* = A_D^* \tilde{c}_1 c_D^* + A_B'^* \tilde{c}_1 c_B'^* \approx A_D^* \tilde{\rho}_{1D} + A_B'^* \tilde{\rho}_{1B} \\
\tilde{\rho}_{24} & \approx \tilde{c}_2\tilde{c}_4^* = A_D^* \tilde{c}_2 c_D^* + A_B'^* \tilde{c}_2 c_B'^* \approx A_D^* \tilde{\rho}_{2D} + A_B'^* \tilde{\rho}_{2B}
\end{split}
\end{equation}

After some simplifications one gets the following equations, which is the relation given in eq.~\ref{approximated coherent terms 4} of the main text.

\begin{equation}\label{4 state 12-coherence2}
\begin{split}
\tilde{\rho}_{14} =\frac{\Omega_4}{2\delta_4 - i\Gamma_4}
\{ (-\cos{\theta_3} \sin{\theta_4} + \sin{\theta_3} \cos{\theta_4} e^{-i\Phi_0})\times \\
 (\cos{\theta_3} ~\rho_{DD} + \sin{\theta_3}~\rho_{BD}) - 
 (\sin{\theta_3} \sin{\theta_4} + \cos{\theta_3} \cos{\theta_4} e^{-i\Phi_0})\times \\
(\cos{\theta_3}~\rho_{DB} + \sin{\theta_3}~\rho_{BB})
\}
\}\\
\tilde{\rho}_{24} =\frac{\Omega_4}{2\delta_4 - i\Gamma_4} 
\{ (-\cos{\theta_3} \sin{\theta_4} + \sin{\theta_3} \cos{\theta_4} e^{-i\Phi_0})\times \\
 (\cos{\theta_3}~\rho_{BD} - \sin{\theta_3} ~\rho_{DD}) - 
 (\sin{\theta_3} \sin{\theta_4} + \cos{\theta_3} \cos{\theta_4} e^{-i\Phi_0})\times \\
 (\cos{\theta_3} ~\rho_{BB} - \sin{\theta_3} ~\rho_{DB})
\}
\end{split}
\end{equation}

\end{document}